# State Space Reduction with Message Inspection in Security Protocol Model Checking

Stylianos Basagiannis    Panagiotis Katsaros    Andrew Pombortsis

*Department of Informatics, Aristotle University of Thessaloniki,*
*54124 Thessaloniki, Greece,*
*{basags, katsaros, apombo}csd.auth.gr*

**Abstract.** Model checking is a widespread automatic formal analysis that has been successful in discovering flaws in security protocols. However existing possibilities for state space explosion still hinder analyses of complex protocols and protocol configurations. Message Inspection, is a technique that delimits the branching of the state space due to the intruder model without excluding possible attacks. In a preliminary simulation, the intruder model tags the eavesdropped messages with specific metadata that enable validation of feasibility of possible attack actions. The Message Inspection algorithm then decides based on these metadata, which attacks will certainly fail according to known security principles. Thus, it is a priori known that i.e. an encryption scheme attack cannot succeed if the intruder does not posses the right key in his knowledge. The simulation terminates with a report of the attack actions that can be safely removed, resulting in a model with a reduced state space.

## 1  Introduction

Security analyses of existing cryptographic protocols have shown that protocols' flaws can be revealed despite the cryptographic primitives used. While perfect encryptions by key-based cryptographic schemes or use of hash functions are considered techniques have been proved secure, the communication procedure may contain logical-based errors that can be exploited by an intruder model. In the related bibliography [1, 2] there are examples of protocols that were published with errors, which remained undiscovered for many years. Thus, formal ways of reasoning [3] for whether a given protocol meets its security goals is an absolute necessity. Model checking is a fully automatic analysis technique that has been successful in discovering flaws in communication protocols. However, ongoing research has not stopped to look for new ways to tackle the problem of state space explosion, which still prevents analyses of complex protocols and protocol configurations (e.g. higher bounds in the number of ongoing protocol sessions).

In general-purpose model checking [4], state space explosion comes from the asynchronous composition of the modeled concurrent processes and the inherent symmetry redundancy of models in many different problem domains. Model checking security guarantees such as secrecy and authentication is based on the hardest possible assumptions for the dominance of the intruder over the communication between the protocol participants. These assumptions represent the general Dolev-Yao intruder model [5]: the intruder can intercept any message transmitted on a public communication channel and can also replace it with a message constructed from his initial knowledge and parts of the messages sent by the participants in the same or in other protocol sessions (intruder's knowledge base). The new messages are created by applying one or more out of four (4) basic operations: encryption, decryption, concatenation and projection. Also, a typical Dolev-Yao intruder model includes additional assumptions, such as the un-breakability of the encryption used and the possibility the intruder to prevent an original message from reaching its destination.

With the mentioned assumptions, any attempt to enumerate all possible attacks in all protocol steps results in an enormous branching of the state space. In the general case, for a given set of eavesdropped messages, the Dolev-Yao operations may be combined recursively, thus producing infinitely many possible fake messages. In explicit state model checking, analysts bound the size of fake messages, in order to set their models finite. However, memory space becomes crucial, due to the need to store information for each state, including the local states of all protocol participants and the accumulated knowledge of the intruder, for the protocol execution.

In current article, we introduce the Message Inspection (MI) intruder model, which is essentially a Dolev-Yao style man-in-the-middle intruder based on the idea of improving his knowledge with protocol-specific metadata that provide information for the exchanged messages. In a preliminary simulation run, the intruder tags the eavesdropped messages with specific metadata parameters enabling him to validate all possible attack actions. The MI algorithm then decides based on this enhanced knowledge, which of the attacks will certainly fail and the simulation run terminates with a report of the attack actions that can be discarded. As a result, it is possible to improve the pruning of the state space by exploiting known security principles, formed as attack actions into the intruder's structure.

We provide a review of the mentioned approaches in Section 2. Section 3 presents the Message Inspection intruder model. The model structure is formally defined and subsequently we introduce the MI algorithm that decides, which attack actions will be performed against the analyzed protocol. In Section 4, we provide experimental results for a MI intruder model in the SPIN model checker [9], when compared with a generic Dolev-Yao intruder model applied upon the Needham Schroeder security protocol (NSPK) [10]. Finally, conclusions are discussed in Section 5.

## 2. Related Work

In related bibliography, there are significant research contributions concerning uses, extensions and improvements of the Dolev-Yao intruder model. Many of these works [2, 11, 12, 13] provided a basis for integrating a custom user-specified intruder model into innovative model checking techniques for the analysis of security properties.

One of the first systems that implemented the Dolev - Yao assumptions and the secrecy failure verification approach was the Interrogator tool [11]. Given a final state in which the intruder knows some message, which should be secret, the Interrogator tries all possible ways of constructing a path that reaches this particular state. If it finds such a path, then it has identified a security flaw. Finite state analysis of cryptographic protocols can take place in specialized security model checkers, like BRUTUS [6], where security violations are encoded as failures of secrecy or authentication. Alternatively, finite state analysis is often carried out in general-purpose model checkers like Murφ [14] and the FDR (Failures Divergence Refinement) [15] model checker. When focused on the problem of the state space explosion, a series of interesting works exploit symmetry and partial order reduction techniques [6, 7, 14, 16]. In [8], the authors propose model checking with pre-configuration, which is a divide-and-conquer method for verifying security protocols.

Athena [17] builds on a different model representation, where in contrast to the conventional trace-based modeling approaches, a set of protocol runs that differ only in the order of interleaving executions of the individual participants is represented by only one state. This is achieved due to a clever extension to the strand space model representation. There is some form of symbolic reduction functionality, but Athena also allows the development of protocol-specific or general pruning theorems. Through this semi-automated approach the analyst uses theorems, in order to prune from the state space all states proved that do not contribute to the final result.

To the best of our knowledge, the MI intruder model is the first model using message metadata collected from a preliminary simulation run. This data enhances the intruder's knowledge with additional information regarding protocol behavior facts that in some cases cannot be observed dynamically across the explored state space paths. It is thus possible to improve the pruning of the state space by exploiting known security principles. Given the intruder's knowledge for the protocol execution, these principles allow determining in advance, whether an attack action can or cannot cause a security violation. We know, for example, that an encryption scheme attack cannot succeed, if the intruder does not possess the right key in his knowledge. Consequently, encrypted messages can be treated differently from plain text or partially encrypted messages and an obvious optimization is to remove from the general Dolev - Yao model all attack actions, for which it is a priori known that they cannot succeed. Attack actions that can be removed are encoded into an open-ended base of primitive attacks (message replays, integrity violations, parallel session attacks and type-flaw attacks) that have been formalized in [18].

## 3   The Message Inspection Intruder Model

This section introduces the MI intruder model and describes its use throughout the preliminary simulation run and the model checking phase. In order to formally describe the proposed approach, we use the formal notation initially specified in [18]. The model can be considered as an optimization approach, where message inspection (MI) reduces the search tree, without excluding any attacks that the analyst needs to check. It is based on a symbolic representation that avoids explicit enumeration of the messages that the intruder can generate from $I_{knowledge}$. Instead of using the Dolev-Yao deduction rules for inferring all possible fake messages in each protocol step, the MI intruder model records the eavesdropped messages in a preliminary simulation run and at the same time creates discrete metadata values for each recorded message. In this way, the intruder model manipulates only the metadata that were initially created and not the messages themselves. MI rules that will be introduced later determine which attack actions are appropriate and must be included in the intruder model for the model checking phase and which are not. Thus, the analyst can use MI, in order to prune the states found to be irrelevant according to the used MI rules.

For the case shown in Figure 1, the MI intruder model acts as a man-in-the-middle attacker that dominates the communication between honest agents *A* and *B*, by eavesdropping the exchanged messages. Each message is evaluated by message characterization mechanisms called *metadata functions*, in order to create appropriate metadata values that enhance the intruder's knowledge for this specific message. The intruder model then consults the embedded base of attack actions, in order to decide which of them can be deactivated by the analyst, without excluding any attacks that may reveal a protocol security flaw. The analyst then proceeds to the model checking phase ($2^{nd}$ protocol execution) with the altered intruder model.

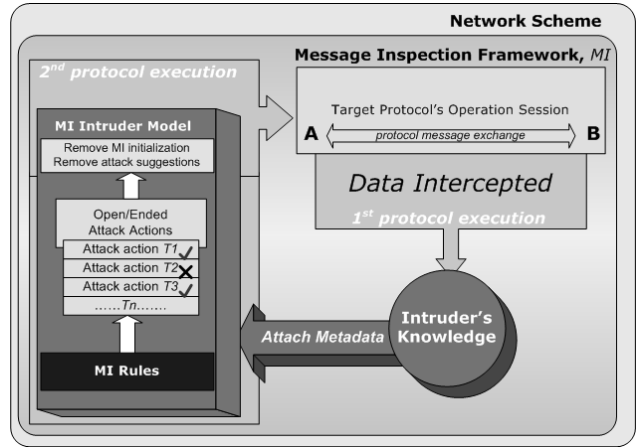

**Fig. 1.** The Message Inspection intruder model

Let us consider a protocol *PR* between participating agents *A, B,...,Z* ∈ `Agents` and let z representing the number of protocol steps. We simulate *PR* for a bounded number of protocol sessions say *n*. We use the messages of the following table, in order to derive metadata for the intruder's knowledge $I_{knowledge}$ as:

$$
\begin{array}{c}
\begin{array}{cccc}
sessions \rightarrow & 1^{st} & 2^{nd} & . & n^{th}
\end{array} \\
\downarrow protocol\ steps
\begin{array}{c}
1^{st} \\
2^{nd} \\
. \\
z^{th}
\end{array}
\begin{bmatrix}
msg_{1,1} & - & - & msg_{1,n} \\
msg_{2,1} & - & - & . \\
. & - & - & . \\
msg_{z,1} & - & - & msg_{z,n}
\end{bmatrix} \\
\underbrace{\qquad\qquad\qquad\qquad}_{\#Ses} \\
\bigcup_{ag\ \in\ Agents}\ \bigcup_{noSes\ =\ 1}^{\#Ses}\ \{sent_{\max(i)}^{ag}\ noSes\}
\end{array} \tag{1}
$$

with $msg_{a,b}$ representing a message sent at the $a^{th}$ step of session $b$ by some ag $\epsilon$ `Agents`. The intruder model stores metadata for each message shown in (1). The stored metadata values for some message sent at the $a^{th}$ step of session b, are derived by a parametric metadata function $p(a,b)$ that is defined as follows.

**Definition 1.** *$p(a,b)$ is a $K^{th}$ parametric metadata function with K sub-functions,*

$$p(a,b) = \begin{cases} p(msg_{a,b})^1 \\ p(msg_{a,b})^2 \\ . \\ . \\ p(msg_{a,b})^K \end{cases}, K \geq 1 \qquad (2)$$

*where the value of $p(msg_{a,b})^{mtd}$, $1 \leq mtd \leq K$ depends on the metadata attribute mtd being expressed (e.g. Encryption, Size etc.) for the specific message $msg_{a,b} \in Msgs$ that is sent at the $a^{th}$ step of session b.*

**Definition 1.1**. *The sub-function $p(msg_{a,b})^{Encryption}$ of $p(msg_{a,b})$ represents the readability of the intercepted message. The image of $p(msg_{a,b})^{Encryption}$ is the set E={0, 1, 2}, where each value denotes one distinct case of encryption form: 0 is used for no encryption, 1 for partial encryption and 2 for a fully encrypted message,*

$$p(a,b) = p(msg_{a,b})^{Encryption} = \begin{cases} 0 & when\ msg_{a,b} = msg_u \\ 1 & when\ msg_{a,b} = msg_y \cdot \{msg_s\}_k \cdot msg_z, \forall (a,b) \in [1..z] \times [1..n] \\ 2 & when\ msg_{a,b} = \{msg_u\}_k \end{cases} \qquad (3)$$

*for some $msg_u \in Msgs$, $k \in Keys$ and $msg_y \cdot msg_z \neq ( )$, i.e. at least one of the concatenated messages is not null.*

Based on the implemented MI function, the value of $p(msg_{a,b})^{mtd}$ may represent e.g. the size of the message or whether the message is readable (plain text) or not. The following definitions instantiate MI for the specific metadata cases of the MI intruder model used in the model checking of the NSPK.

MI enables the intruder model to act as a decision-making machine that groups attack actions into three different operational procedures corresponding to the symbolic values 0 for no encryption, 1 for partial encryption and 2 for full encryption. In this way, the MI intruder model implements the additional capability to select attack actions, for which according to known security principles – encoded as MI rules in Table 2 – it is a priori known that they will not succeed. For example, an encryption scheme attack will not uncover a protocol flaw, if the intruder does not possess the right key in $I_{knowledge}$. Instead of model checking a series of meaningless attacks, the MI algorithm informs the analyst for the possibility to correct his model by removing them. Thus the intruder model is simplified and in effect performs only the necessary attacks. Each attack action belongs to one of the broad categories of attacks, which were formalized in [18] as specific sequences of "send" and "receive" actions.

**Definition 1.2**. *The sub-function $p(msg_{a,b})^{Size}$ of $p(a,b)$ represents the message size in bits for some intercepted message. The image of this sub-function is some set of symbolic values S={s: s $\epsilon$ $\mathbb{N}$ and s>0} with natural numbers representing valuations of the size of messages for the modelled protocol.*

$$p(a,b) = p(msg_{a,b})^{Size} = \begin{cases} size(msg_{a,b}) & \text{for some mapping } size \subseteq Msgs \times S \\ 0 & \text{if } msg_{a,b} \text{ is never sent (null)} \end{cases}, \forall (a,b) \in [1..z] \times [1..n] \qquad (4)$$

This specific sub-function enables the MI intruder model to track a message as a numeric valuation of its size, which in turn depends on the size valuations of its constituent parts. When the MI intruder detects two metadata values in different protocol sessions that correspond to messages of equal size, then according to [19] it is possible to mount a type flaw attack, irrespective of whether this attack will succeed or not. Furthermore, if the protocol is interrupted by some communication error and the intruder model stops receiving messages (timeout), then the size of the expected messages in all future steps of the same protocol session will be zero (0). In this case, the intruder model ignores the metadata values of this particular sub-function for the undelivered messages.

The columns of the table shown in (1) represent numbered steps in the simulated protocol sessions. These columns are seen as monotonically increasing sequences with positive integer terms $b_m \in \aleph$ and $b_0 = 1$. The different terms can be considered as *message timestamps* that are set by the intruder model for the intercepted messages. They imply a relative message ordering that for two messages taken from the same or from interdependent protocol sessions may be used for checking whether one message precedes the other or not. As noted in related work, the applicability of some attack actions depends on the availability of intercepted message parts with timestamp values that are related in some way to the timestamp of the last intercepted message in the attacked protocol session. For example, an impersonation attack between two parallel sessions cannot – according to [19] and [20] – reuse message parts, with timestamp values greater than the timestamp value of the last intercepted message in the attacked protocol session. If necessary, the MI intruder model can integrate additional metadata sub-functions besides those mentioned. After having defined all metadata sub-functions, we define now the *Intruder Knowledge Table [Ikt]* as follows:

**Definition 2.** *In a MI intruder model we define the intruder knowledge table [Ikt], which is populated with the values of a parametric metadata function p(a,b) for all intercepted messages* $msg_{a,b}$, *with (a,b)* $\in [1..z] \times [1..n]$:

$$[Ikt] = \begin{bmatrix} p(1,1) & . & . & p(1,n) \\ p(2,1) & . & . & . \\ . & . & . & . \\ p(z,1) & . & . & p(z,n) \end{bmatrix}$$

*with*

$$p(a,b) = \begin{cases} p(msg_{a,b})^{Size} = \begin{cases} size(msg_{a,b}) & \text{for a mapping } size \subseteq Msgs \times S \\ 0 & \text{if } msg_{a,b} \text{ is never sent (null)} \end{cases} \quad \forall (a,b) \in [1..z] \times [1..n] \\ \\ p(msg_{a,b})^{Encryption} = \begin{cases} 0 & \text{when } msg_{a,b} = msg_u \\ 1 & \text{when } msg_{a,b} = msg_y \cdot \{msg_u\}_k \cdot msg_z \\ 2 & \text{when } msg_{a,b} = \{msg_u\}_k \end{cases} \\ \\ \text{additional sub-functions ...} \end{cases} \qquad (5)$$

*for some set of symbolic values S={s: s ∈ ℕ and s>0} with natural numbers and some $msg_u$ ∈Msgs, k ∈Keys and $msg_y \cdot msg_z \neq ($ ). The properties of [Ikt] are:*

- *if $msg_{a,b}$ is never sent (null) then p(a,b)=0 and this means that the intruder has not intercepted any message sent in the $a^{th}$ protocol step of session b*
- *if p(a,b)=0 then p(a+φ, b)=0 $\forall \varphi \in$ ℕ: a+φ≤z*

The properties of the [Ikt] table enable manipulation of the collected metadata values, for deriving protocol-specific model checking improvements like for example a state space reduction, through simplification of the applied intruder model. For a protocol PR and an intercepted message in protocol step a of session b the intruder model fills in the metadata values p(a, b). If a < z, then all table entries p(a+φ, b) with φ∈ ℕ: a+φ≤z keep their initial value, which is zero (0), until the intruder intercepts the respective message. If for some reason, the protocol session is stopped, then the values of p(a+φ, b) remain zero.

**Definition 3.** *In order to compare two different [Ikt] table entries, say p(a, b) and p(c, d), such that a≠c ∨ b≠d, we define the following operator: p(a, b) ≅ p(c, d), iff*
$$\left(p(msg_{a,b})^1 = p(msg_{c,d})^1 \vee p(msg_{a,b})^2 = p(msg_{c,d})^2 \vee ... \vee p(msg_{a,b})^K = p(msg_{c,d})^K\right)$$

In the preliminary simulation run, the intruder's knowledge is enhanced with the metadata of the [Ikt] table. The intruder's knowledge then includes the information
$$I_{knowledge} = \bigcup_{ag \in Agents} \bigcup_{noSes=1}^{\#Ses_{ag}} \{sent_{\max(i)}^{ag_{noSes}}\} \cup I_{in\_knowledge} \cup \{[Ikt]\}.$$ In this first phase, the intruder model acts as a passive model entity, i.e. it does not execute "send" actions against honest participants. The performed simulation applies the MI algorithm to the updated $I_{knowledge}$ and enables the intruder model to manipulate the message sequences $\bigcup_{ag \in Agents} \bigcup_{noSes=1}^{\#Ses_{ag}} \{sent_{\max(i)}^{ag_{noSes}}\}$ for all ag∈Agents based on the [Ikt] table. The obtained simulation output may include a list of attack actions that can be safely removed from the MI intruder model.

Figure 2 introduces the two phases of the MI algorithm. We consider agents A, B ∈ Agents that exchange messages according to process descriptions $P_I$ and $P_R$ with the actions performed in the roles of the initiator and the responder for some protocol, say PR. The intruder model acts as a man-in-the-middle entity that captures all messages exchanged between protocol participants. For each intercepted message, the intruder model creates a structure p(a, b) corresponding to the [Ikt] table entry for the $a^{th}$ step of session b, as shown in Figure 2. The metadata values in the p(a, b) structures are used for comparing the intercepted protocol messages, in order to select the applicable attack actions.

Let us consider that protocol PR has four (4) steps and runs in two sessions. In the *MI initialization phase*, the intruder model records all intercepted messages from the two protocol sessions. The number of fields in the created structures p(a,b) is the number of metadata sub-functions that are implemented. When the MI intruder intercepts a message, it updates the respective [Ikt] table entry, which is used for comparing it – by applying operator ≅ as defined in definition 3 – with the table entries of previous rows or if the message is part of the second session, with the table entries of the first column. Each comparison determines if there are attack actions that according to the MI rules of Table 2 can be safely excluded, in the examined protocol step. Attack actions that do not contribute in the model checking for all protocol steps are reported in the produced simulation output and it is then possible to remove them from the MI intruder model. The analyst also removes the MI initialization part and proceeds to the model checking of the security guarantees of interest, with the optimized MI intruder model that generates a reduced state space.

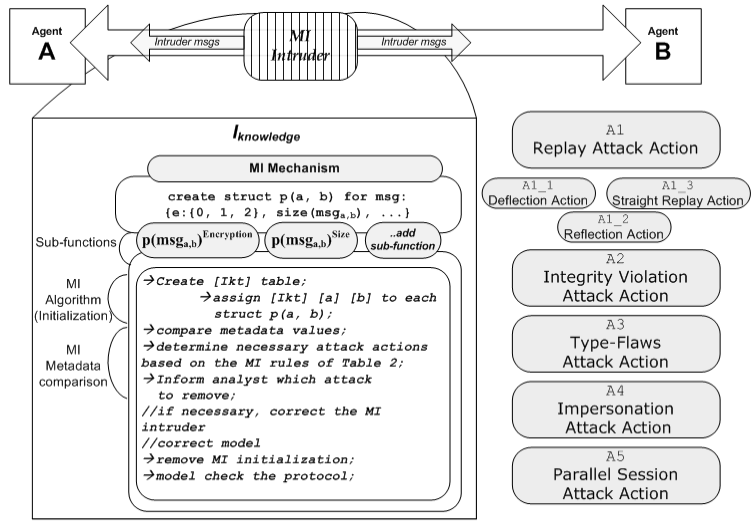

**Fig. 2.** The MI algorithm and the attack actions for the MI intruder model

Figure 2 also shows the open-ended base of attack actions that in the current implementation of the MI intruder model are checked for their feasibility.

**Table 1.** Attack actions of the MI intruder model and how they are related to the meta-data entries of the [Ikt] table

| Attack Action | Action description |
|---|---|
| **A1** | *Select an intercepted message and send it to its sender (A1_2) or to its intended recipient (A1_3) or to some participant that is neither the intended recipient nor the sender (A1_1)* |
| **A2** | *Replace an intercepted message with another message or produce a fake message by concatenation with some message from $I_{knowledge}$* |
| **A3** | *Replace a (part of an) intercepted message that corresponds to some p(a,b) with a previously intercepted message (part)* |
| **A4** | *Impersonate some ag $\in$ Agents using a previously intercepted message that corresponds to some p(a,b) with a=1* |
| **A5** | *Initiate a new protocol session or manipulate an existing session using a previously intercepted message that corresponds to some p(a,b)* |

The selected attack actions appear as primitive steps in attacks reported in the related bibliography and have been proposed in published taxonomies [20, 21] that formalize the observations of intruder misbehaviors, where the intruder redirects messages among protocol participants. In [18], we provided formal definitions of the selected primitive attack actions, as well as bibliographic examples, where these attack actions violate security properties of existing protocols.

Table 1 introduces textual descriptions of the sequences of "send" and "receive" actions for the attack actions of Figure 3, as well as how these actions are related to the meta-data entries of the [*Ikt*] table. Attack actions A1 represent the sending of an intercepted or (if combined with another attack action) a counterfeited message, either to its original sender or to its intended recipient or even to some participant that is neither the intended recipient nor the sender. The metadata values $p(a, b)$ do not influence the feasibility of this general attack action. However, we adopt the assumption that if the sent fake message does no comply with the pattern of the message expected by the "victim", then the recipient falls into a fail-stop state, i.e. he does not continue with the ongoing protocol execution. This assumption represents the expected behavior of a correct protocol implementation.

Attack action A2, when feasible, alters an intercepted message by replacing it or part of it with some message from $I_{knowledge}$. This is possible only when $p(msg_{a,b})^{Encryption}$ is 0 or 1. If $p(msg_{a,b})^{Encryption}=2$ and the intruder does not have in $I_{knowledge}$ the right key for decrypting the intercepted message, then the contents of the message cannot be read (un-breakability of the encryption used) and the A2 attack action is not possible.

**Table 2.** Rules for checking feasibility of attack actions for MI intruder model

| Metadata | Enabling conditions | | Attack Actions |
|---|---|---|---|
| Readability $p(msg_{a,b})^{Encryption}$ | $p(msg_{a,b})^{Encryption}= 2$ | | A1, A4, A5 |
| | $p(msg_{a,b})^{Encryption}= 1$ | | A1, A2, A4, A5 |
| | $p(msg_{a,b})^{Encryption} \neq 2$ and $\exists\, m \in Msgs: exists(m, msg_{a,b}) = \texttt{true}$ and $\exists\, amsg \in AMsgs \cap I_{knowledge}:$ $p(amsg)^{Size}=p(m)^{Size}$ | | A3 |
| | $p(msg_{a,b})^{Encryption}= 0$ | | A1, A2, A4, A5 |
| Size $s_1 = p(msg_{a,b})^{Size}$ and $s_2 = p(msg_{c,d})^{Size}$ | $s_1 = s_2$ and $a < c$ | $b = d$ | A3 |

Attack action A3 replaces a part of an intercepted message or the whole message, with another message from $I_{knowledge}$. The produced fake message can be accepted by the "victim", only if its size is the same with the size of the expected message. This can be checked by appropriate comparisons of stored metadata values for the messages in $I_{knowledge}$. Type flaws with partially altered messages are possible only when $p(msg_{a,b})^{Encryption}$ is not 2, i.e. when the intercepted message is (partially) readable. Alternatively, according to [20], a type flaw attack is also possible, when in a protocol session an honest agent falls into misinterpretation of a received message, supposed to deliver specific data in some protocol step. This type flaw attack is an open possibility even when the used intercepted message is fully encrypted and the intruder does not possess in $I_{knowledge}$ the key needed to decrypt it.

Attack action A4 initiates a new protocol session be reusing a previously intercepted message that corresponds to some $p(a,b)$ with $a = 1$. Finally, attack action A5 initiates a new protocol session or manipulates an existing session by reusing a previously intercepted message. Both A4 and A5 are not based on specific requirements for the encryption form of the intercepted message.

Table 2 introduces the MI rules for checking feasibility of attack actions in the first implementation of the MI intruder model. The enabling conditions are used in metadata comparisons like the ones described in next paragraphs, in order to determine whether an attack action is feasible or not. Attack actions that in all protocol steps are not feasible can be safely removed, thus yielding an optimized intruder model for the analyzed protocol. The metadata sub-function $p(msg_{a,b})^{Encryption}$ plays an important role in this analysis, since its values determine whether the intercepted message $msg_{a,b}$ can or cannot be read. When $msg_{a,b}$ is fully encrypted ($p(msg_{a,b})^{Encryption}=2$), the intruder model checks in $I_{knowledge}$ if it owns the key needed to decrypt the intercepted message. If the key is found, this message is marked as non-encrypted and the metadata value $p(msg_{a,b})^{Encryption}=0$ is recorded in the corresponding field of $p(a,b)$. If it is possible to read only some part of the intercepted message $msg_{a,b}$ then $p(msg_{a,b})^{Encryption}=1$, i.e. $msg_{a,b}$ is partially encrypted. This is a sufficient condition for enabling attack actions A1, A2, A4 and A5. Moreover, the possibility to replace a part of the message, say $m$, with some atomic message $amsg$ from $I_{knowledge}$ requires equal metadata values for $p(amsg)^{Size}$ and $p(m)^{Size}$. This enabling condition implements the requirement for making an agent vulnerable to misinterpret some part of the message (attack action A3), which is its size. In most cases, $p(msg_{a,b})^{Encryption}$ will be 2, which automatically excludes the possibility of an integrity violation (attack action A2) that requires read access to some part of the intercepted message. As we already noted, in type flaw attacks where the intercepted message is replaced as a whole, there is no special requirement for its encryption form. If the expected message has the same size with an intercepted message from a previous step of the same protocol session [20], then it is possible for the intruder to mount a type flaw attack. In the last row of Table 2 we provide the enabling conditions for this attack action. For a complete description of the attack actions mentioned in Table 2 the reader is referred to [18].

For all attack actions of Table 1, the intruder model compares the metadata values of the intercepted messages and if an attack action is possible, the model indicates the feasibility of the examined attack action.

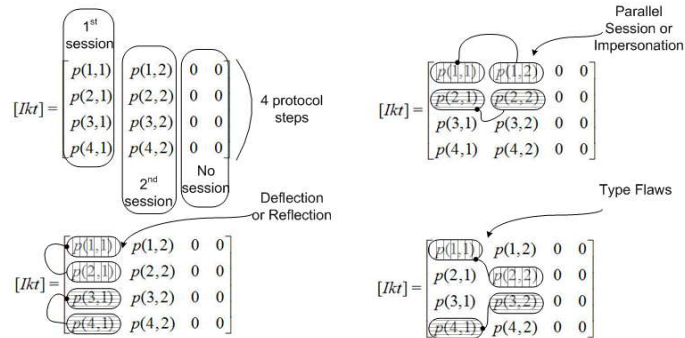

**Fig. 4:** Comparisons of the [Ikt] table entries for detecting possible attack actions

Examples of the comparisons made for the considered 4-step protocol are shown in Figure 4, for most of the mentioned attack actions. When $p(1,1) \cong p(1,2)$ and at the same time holds for these two table entries one of the conditions of Table 2 that enable attack actions A4 and/or A5, then the MI intruder model can initiate a new protocol session, in order to attempt a parallel session or an impersonation attack. When $p(2,1) \cong p(2,2)$ and for these two table entries hold the conditions of Table 2 that enable attack action A5, then it is possible the MI intruder to manipulate an existing parallel session for an attack, that may subvert one of the protocol's correctness properties. When $p(a,b) \cong p(c,b)$ and for these two table entries hold the conditions of Table 2 that enable attack actions A1, then it is possible the MI intruder to perform a deflection or a reflection message replay. Finally, when $p(a,b) \cong p(c,d)$ for two messages in different protocol sessions, the last intercepted message is (partially) readable and at the same time hold the conditions shown in Table 2, then it is possible the MI intruder to perform a type-flaw attack action.

In that case $I$ trigger the attack, possibly after having altered the eavesdropped $msg \in Msgs$ based on $I_{knowledge}$, thus resulting in a $msg' \in Msgs$. The subsequent action performed by $I$ is either $send\ (I, v, msg')$ or $send\ (I, v, \{msg'\}_k)$ for a $k' \in I_{knowledge}$ such that $v \in is\_key\_of\ (k')$, i.e. $v$ is the owner of $k'$. This attack action succeeds, if in the global state after the occurrence of the action $receive\ (v, I, msg')$ or respectively $receive\ (v, I, \{msg'\}_k)$ there is some atomic message $amsg$, such that $exists(amsg,\ rcvd_{\max(i)}^{v_{noSes}}) = \mathtt{true}$, $1 \leq noSes \leq \#Ses_v$ and for two sets $Set_e$ and $Set_f$ from the "disjoint" union $Amsgs$, $amsg \in Set_e \cap Set_f$ where $i \geq 1$ represent the terms of the concatenation sequence of messages received by agent $v$ in the course of session $noSes$. Thus, an atomic message that was originally intended to have one type (e.g. nonce) is interpreted as having another type (e.g. key or data) meaning that the type flaw is exploited, even though this may not lead to a direct security compromise.

During model checking, the MI intruder model performs all possible attack actions in all protocol steps, after having excluded - as a result of the preliminary simulation run - the attack actions whose enabling conditions are not satisfied in all protocol steps. The honest agents $\mathtt{ag} \in Agents$ either accept or reject the fake messages based on the implemented protocol logic. In essence, the metadata in the [$Ikt$] table store protocol-specific monitoring information, which is used in controlling the behavior of the MI intruder model in an effective way.

## 5 MI-based model checking of the NSPK protocol

The NSPK protocol aims to establish mutual authentication between the initiator and the responder, in order to start a message exchange between them. The protocol name suggests the use of public key cryptography, for delivering authentication guarantees. The reduced version of the NSPK protocol, shown in Figure 6, includes only three (3) protocol steps, where in each step the protocol participants $A$ (for the initiator) and $B$ (for the responder) exchange messages with agent identities and randomly generated nonces ($N_A$, $N_B$), encrypted by the public keys $PK\{A\}$ and $PK\{B\}$. The sent information can be checked by the receivers.

The reported results concern with two different intruder models, i.e. the general Dolev-Yao intruder with the deduction rules specified in [5] and an MI intruder model. Honest agents are encoded as fail-stop processes, i.e. if the message received in a protocol step is not the expected one, then protocol execution is stopped for the receiver even though the provided security guarantees have not been violated. Potential failures of message secrecy or failures of authentication are expressed as invalid end-states in the SPIN model checking environment.

In a preliminary simulation run with two protocol sessions (Figure 6), the MI intruder model detected two attack actions, namely A2 and A3 that can be safely removed. More specifically, the MI intruder acts as a man-in-the-middle entity between agents A and B for the first protocol session and B and C for the second session. Upon intercepting an NSPK message, say $msg_{a,b}$, the MI model creates appropriate metadata values for $p(msg_{a,b})^{Encryption}$ and $p(msg_{a,b})^{Size}$ that are recorded in [$Ikt$] table (Figure 6i). Since the MI intruder forwards the sent messages to the intended recipients, both protocol sessions are completed with success. We realize that all protocol messages are fully encrypted and since the decryption key is never included in $I_{knowledge}$, for all metadata values $p(msg_{a,b})^{Encryption}$ is 2. We also see the metadata values computed for $p(msg_{a,b})^{Size}$, for the message sizes shown in Figure 6, representing the assumption that there are no two size-similar atomic messages, with the first message being an agent identity and the second one coming from the set of nonces.

The MI intruder model then performs (Figure 6ii) the metadata comparisons discussed in section 4.2 by taking into account the MI rules of Table 2. Finally, the intruder model outputs the decisions made (Figure 6iii). Since $p(msg_{a,b})^{Encryption} = 2$ in all protocol steps, the integrity violation attack action (A2) is excluded. Also, because $p(msg_{a,b})^{Encryption} = 2$ for all exchanged messages and at the same time there are no size-similar messages in the same protocol session, the MI-based intruder model proposes removing the type flaw attack action (A3).

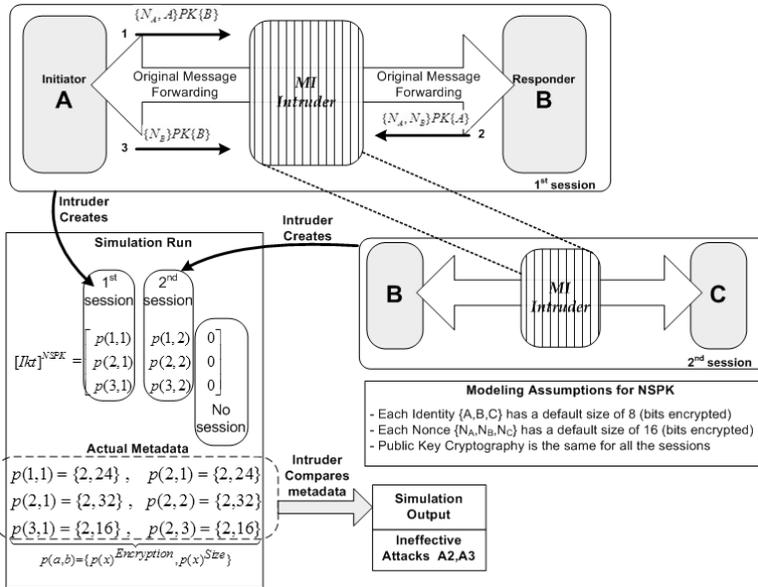

**Fig. 6.** Preliminary MI simulation run: the intruder (i) creates the [Ikt] table, (ii) compares the metadata and (iii) proposes removal of attack actions A2 and A3

Figure 7a provides the model checking result for the described NSPK model, when using the optimized MI intruder and the partial order reduction functionality of the SPIN model checker. An invalid end state is reached at depth 25 of the produced reachability graph and the search is stopped after having detected the reported error. A subsequent guided simulation explores the generated counterexample and creates the message sequence diagram of Figure 7b. The reached invalid end state corresponds to the state, where agent B acting as responder accepts a fake NSPK message that causes him to initiate a new protocol session. The used message is created by the intruder model in the role of the initiator according to the message pattern of the first of the three messages shown in Figure 6. This invalid authentication in effect causes a successful impersonation attack against B, who perceives the intruder as an honest protocol participant. The same security violation has been also detected with a generic Dolev-Yao intruder model, but in that case the reached depth of the detected invalid end state was 48. The compared objects are the state spaces produced for the two intruder models where we compare the size of the spaces generated up to the point of having detected the error (shown in Figure 8).

Obviously, the size reduction achieved by the optimized MI intruder model in this case is orders of magnitude less than the size reduction achieved in the aforementioned more general case. The default working mode of the SPIN model checker includes the partial order and symmetry reduction functionality. However, in order to explore the interactions of the two intruder models with additional reduction and state exploration options of SPIN, we also report search space techniques produced in other error exploration modes (DFS and BFS search modes).

```
pan: invalid end state (at depth 25)
pan: wrote pan_in.trail

(Spin Version 5.1.6 -- 9 May 2008)
Warning: Search not completed
        + Partial Order Reduction

Full statespace search for:
    never claim          - (not selected)
    assertion violations - (disabled by -A flag)
    cycle checks         - (disabled by -DSAFETY)
    invalid end states   +

State-vector 162 byte, depth reached 24, errors: 1
      264 states, stored
      883 states, matched
     1147 transitions (= stored+matched)
       11 atomic steps
hash conflicts:   1493 (resolved)

    2.326       memory usage (Mbyte)

unreached in proctype Agent_A
    line 33, "pan_in", state 8, "-end-"
    (1 of 12 states)
unreached in proctype Agent_B
    (0 of 10 states)
unreached in proctype MI_intruder
    line 54, "pan_in", state 8, "-end-"
    (1 of 86 states)
```

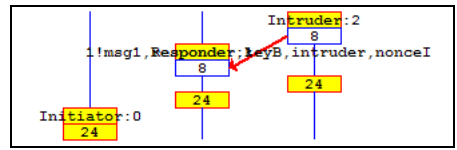

**Fig. 7a.** Verification output for NSPK with the detected invalid end state at depth 25

**Fig. 7b.** Guided simulation of counterexample with the detected impersonation

While SPIN uses by default partial order reduction technique [9] among with a DFS search, we can disable the specific feature or alter the search algorithm (from DFS to BFS search). The main advantage of the BFS search option - which is effective only for safety properties (secrecy and authentication) - is that it finds the shortest path to an error state, while the DFS search often finds a longer path. Figure 8, provides results for the state spaces generated up to the point of having detected the invalid end state. The NSPK protocol model with the Dolev-Yao intruder model generated state spaces with about 2.5 times the number of stored unique states for the MI intruder. An interesting observation is that this improvement depends on the depth where the error is discovered and from this point of view the Breadth-First Search finds the shortest path to the error.

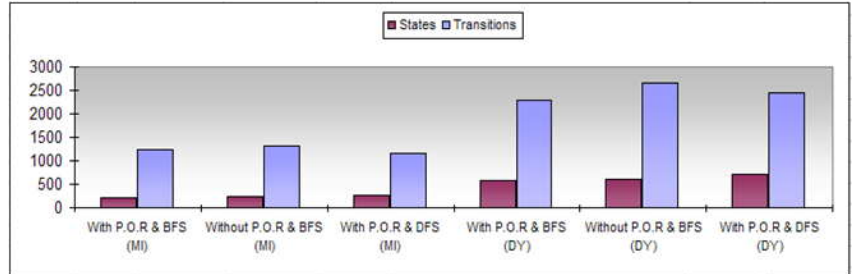

**Fig. 8.** Size of the state spaces for the NSPK security flaw with the optimized MI intruder model (MI) and the general Dolev-Yao intruder model (DY)

While in Figure 7a (MI intruder with Depth-First Search) the error was detected at depth 25 ($1.4 \cdot 10^3$ hash collisions in a hash table with 264 stored unique states), when using Breadth-First Search the error was discovered at depth 5 (no hash collisions in a hash table with 112 unique states). On the other hand, when verifying NSPK with the Dolev-Yao intruder model and Depth-First Search the error is detected at depth 48 resulting in $7.3 \cdot 10^4$ collisions in a table with 698 stored states.

## 6. Conclusion

The MI intruder model aims to restrict the inherent combinatorial complexity of security model checking with intruder models that adopt the general Dolev-Yao rules. This is achieved through message inspection that allows customizing the intruder behavior, by taking into account protocol specific metadata for the structure and the characteristics of the exchanged messages. The described modeling approach does not exclude other state space pruning techniques. The only requirement is that it can be implemented only in model checking environments that support both simulation and model checking of the analyzed security protocol. We conducted a series of experiments for exploring the improvements in the model checking of the NSPK protocol, when compared with the generic Dolev-Yao intruder model.

The MI intruder model provides an open-ended framework for integrating additional protocol-specific model checking optimizations. For a potential extension concerning insertion of feasibility check for a new attack action the analyst will have to implement additional MI rules and metadata comparisons. If necessary, the model may be extended by including additional metadata parameters, but this will cause modifications in the MI initialization code that stores metadata values for the intercepted messages in the [*Ikt*] table. A future goal is the design of an integrated modeling environment that will provide the described functionality in a usable model checking package.